\documentstyle[11pt,paspconf]{article} 

\newcommand{\avg}[1]{{\langle{#1}\rangle}}

\begin{document} 

\title{Constraining the Dark Matter Distribution With Large-Scale
Structure Observations}
\author{Michael A.~Strauss and Michael Blanton}
\affil{Princeton University Observatory, Princeton, NJ 08544}

\begin{abstract}
We discuss the use of galaxies to trace the large-scale structure of
the universe and thereby to make cosmological inferences. We put special
emphasis on our lack of knowledge about the relative distribution of
galaxies and the dynamically important dark matter. We end with a
discussion of the increasing importance of infrared astronomy to
large-scale structure studies.
\end{abstract}


\section{{\bf Introduction}}

Modern observational cosmology is done within a standard paradigm that
has been in development since the invention of the concept of the
expanding universe. It is worth reminding ourselves what the basic
tenets of the paradigm are.
\begin{itemize}
\item We live in a uniformly expanding universe, which originated in a
Hot Big Bang. The evidence for the uniform expansion is overwhelming:
independent measurements of redshift and distance show that the two are
proportional to impressive accuracy. This has been most spectacularly
demonstrated recently with observations of high-redshift supernovae
({\it e.g.}, Schmidt~{\it et~al.} 1998), which are found to obey the
Hubble Law from $z=0.003$ to $z\approx 1$\footnote{Of course, the
supernovae are showing evidence of deviations from the linear
relation, which is interpreted to be due to the curvature of the
Universe, but that is a different story.}.
\item The universe is homogeneous and isotropic on large scales, and
thus is described in General Relativity by the
Friedman-Robertson-Walker (FRW) metric. This is a statement of the
Cosmological Principle, and is dramatically demonstrated by the
isotropy of the Cosmic Microwave Background (CMB) to one part in
$10^5$.
\item The large-scale distribution of matter grew via gravitational
instability from initially low-amplitude fluctuations. The CMB indeed
does show deviations from isotropy, which are interpreted to be due to
tiny fluctuations in the initial density field. Gravity amplifies the
contrast between overdensities and underdensities, eventually leading
to the structures that we see today.
\item Dark matter dominates the mass density of the universe. There is
strong evidence that most of the mass density of the universe is in a
form that is not directly visible, manifesting itself only through its
gravitational influence, and that the baryonic material makes up of order
10\% or less of the mass density of the universe.
\end{itemize}
This paradigm leaves many important questions unanswered.
\begin{itemize}
\item The FRW metric is described by a number of parameters (not all
independent), which we might hope to measure: $\Omega$, the mass
density of the universe relative to the critical density; $H_0$, the
Hubble Constant; $\Lambda$, the contribution to the curvature by
vacuum energy; $q_0$, the deceleration parameter; and $t_0$, the
current age of the universe. The quantity $\Omega$ can be further
divided into the contributions from different mass constituents:
baryons, and hot and cold dark matter. Our model will not be complete until
we have estimates of these parameters, and show that they are mutually
consistent within the model.
\item We must characterize the initial perturbations from which the
present-day large-scale structure grew. To the extent that the Fourier
modes of the initial perturbations had random phases, as expected in
inflationary models, the {\it power spectrum} gives a complete
statistical description of the perturbations. If the fluctuations were
seeded by discrete structures, as one expects in models involving
phase transitions in the early universe, then the phases are
correlated, and one needs higher-order statistics to describe the
density field.
\item The physical nature of the dark matter remains unknown. We have
good indirect evidence that most of it is non-baryonic, and because it
is dark, we know that it does not interact strongly with photons. Its
detailed nature will affect the shape of the power spectrum of
fluctuations, so measurements of the power spectrum can shed light on
the nature of dark matter.
\item Dark matter may dominate the mass density of the universe, but
it is galaxies that we see. A complete model must describe how and
when they formed, and how their formation is tied to the distribution
of dark matter. Our theories can most easily predict the statistics of
the distribution of dark matter; the relative distribution of galaxies
and dark matter remains an important unknown in interpreting the
observed large-scale structure of galaxies.
\end{itemize}

In this paper, we do not attempt to answer all these rather broad and
important questions; rather, we discuss how to get a handle on all of
them by measuring the large-scale distribution and motions of
galaxies, with special emphasis on the last problem mentioned above:
the relative distribution of galaxies and dark matter. We will end
with a brief description of the impact of infrared astronomy on the
study of large-scale structure.

\section{Large-Scale Structure Data}

Imperfect though they may be, galaxies are the tracers that we use to
learn about the distribution of mass on large scales.

The NGC was the first substantial catalog of galaxies compiled; it was
apparent even from maps of these few thousand objects on the sky that
they are not distributed uniformly. In particular, the plane of what
we now call the Local Supercluster is quite apparent in the NGC, with
a particular concentration in the Virgo Cluster. Although we can learn
a great deal from the projected distribution of galaxies on the sky,
including the power spectrum and other clustering statistics, we need
to know the distance of the objects in a galaxy catalog to quantify
all the properties of the large-scale distribution of galaxies. The
Hubble Law, which relates the redshift of a galaxy to its distance,
allows us to tease out the all-important third dimension of the galaxy
distribution. Since the first substantial redshift surveys for the
study of large-scale structure in the late 1970s, surveys have grown
in size and completeness; currently, the largest single redshift
survey is that of Schechtman {\it et al.}~(1996), which contains
roughly 25,000 galaxies.

Even at low redshift, where cosmological curvature effects are
unimportant, the Hubble Law holds exactly only in the case of a
perfectly homogeneous universe. The over- and under-densities we are
concerned with here exert a gravitational pull on the adjacent
galaxies. This gives galaxies a {\it peculiar velocity}, above and
beyond the pure Hubble flow. The radial component of these velocities
enters into the redshift:
\begin{equation}
cz = H_0 r + {\hat {\bf r}} \cdot [ {\bf v} ({\bf r}) - {\bf v}({\bf
0})] \mathrm{.}
\end{equation}

One can thus estimate the radial component of these peculiar
velocities using independent measurements of the redshifts $cz$, and
distances, $H_0r$, of galaxies. Of course, the measurement of
distances is inherently difficult, especially for extragalactic
objects; the standard candles one uses (such as the Tully-Fisher
relation, which allows one to estimate the absolute magnitude of a
spiral galaxy, given its rotation speed) have substantial
scatter. Nevertheless, peculiar velocity studies have been done for
large samples of galaxies, which are starting to yield a picture of
the large-scale velocity field.

With these redshift and peculiar velocity data, one can carry out a
number of studies:
\begin{itemize}
\item We can ask for the large-scale topology of the galaxy
distribution. Even if the initial density field satisfied the
random-phase hypothesis, nonlinear gravitational growth will cause
coherent structures to form. People have described the galaxy
distribution as being in sheets, filaments, the walls of bubbles, and
so on. Interestingly, there still is not a satisfying statistical
method of describing the features that are so apparent to our
pattern-seeking eyes.
\item We can calculate the power spectrum of the galaxy
distribution. This has been distorted in various complicated ways from
the initial power spectrum of the early universe due to nonlinear
growth, the effects of peculiar velocities, and so on, but still has
much to teach us about the initial fluctuations, and by inference, the
nature of dark matter, and the values of cosmological parameters.
\item We can calculate higher-order statistics of the clustering as
well. As we have mentioned several times above, the nonlinear growth
of clustering can take an initially random-phase distribution, and
generate coherent structures, for which the power spectrum does not
give the entire statistical description of the density field. One can
calculate the growth of the higher-order clustering measures with time
using perturbation theory, which can then be checked against
observations; to the extent they agree, one has a consistency check on
the hypotheses of an initially random-phase field and the growth of
structure via gravitational instability.
\item We can relate the large-scale distribution of galaxies to the
peculiar velocities. We believe that it is the gravitational influence
of the density inhomogeneities in the galaxy distribution that causes
the peculiar velocities. In fact, to linear order in the fluctuations,
gravitational instability predicts a simple relationship between the
velocity field ${\bf v}$ and the mass density field $\delta$:
\begin{equation}
\nabla\cdot{\bf v}({\bf r}) = - \Omega^{0.6} \delta({\bf r})\mathrm{.}
\end{equation}
\end{itemize}

The status of all these analyses has been reviewed extensively in the
literature (for example, Strauss \& Willick 1995 and Strauss 1998a,b);
here we want to stress the limitations of existing data and
analyses. First, for many of the analyses that one would like to do in
large-scale structure, existing datasets survey too small a volume. An
obvious example is the power spectrum on large scales. On sufficiently
large scales, one expects the power spectrum to have the same shape as
a function of wavenumber (albeit with a much greater amplitude) as the
initial power spectrum; moreover, the features in the power spectrum
that are diagnostics of the cosmological parameters and the nature of
the dark matter manifest themselves on large scales, above 50 $h^{-1}$
Mpc or so. However, in order to measure the clustering signal on these
scales, one needs a substantial number of independent volumes of this
size. Existing redshift surveys simply do not cover a large enough
volume to get a high signal-to-noise measurement of the shape of the
power spectrum on these scales and larger.

Second, these data are generically affected by systematic errors. Many
of the state-of-the-art large-scale structure studies are based on
galaxy catalogs selected by eye off of photographic plates in the 1970s
and earlier. As the clustering on large scales is weak, it takes only
small inhomogeneities in the selection of the catalog to swamp the
cosmological signal one is looking for.

New and larger redshift surveys, such as the Sloan Digital Sky Survey
(Gunn \& Weinberg 1995; {\it cf.},
{\tt http://www.astro.princeton.edu/BBOOK/}) and the Two-degree Field
redshift survey (Colless 1998; {\it cf.},\\ {\tt
http://mso.anu.edu.au/~colless/2dF/}), will be able to address these
problems. But there is another problem, astrophysical in nature, which
places a more fundamental limitation on the interpretation of large-scale
structure, namely our lack of knowledge of the relative distribution
of galaxies and dark matter.

\section{How are Galaxies Distributed Relative to Dark Matter?}

The physics of dark matter is relatively simple: being collisionless
(as we infer, given its lack of interaction with baryons or photons),
gravitational physics describes its evolution and its large-scale
distribution in the context of any specific model. However, it is
galaxies which we observe directly, and it is not at all obvious {\it
a priori} whether the large-scale structure we observe in the galaxy
distribution mirrors that in the underlying, and presumably
dynamically dominant, dark matter.

Let us define the density field $\rho({\bf r})$ of either galaxies or
dark matter, smoothed on a scale $R$. The Cosmological Principle
states that the universe is homogeneous on large scales; it makes
sense to speak of a mean density $\avg{\rho}$ of the universe. thus we
find it useful to define the density {\it fluctuation} field relative
to this mean density:
\begin{equation}
\delta({\bf r}) = \frac{\rho({\bf r}) -
\avg{\rho}}{\avg{\rho}}\mathrm{.}
\end{equation}

The simplest relation one could imagine between the galaxy and mass
distribution is that they are the same:
\begin{equation}
\delta_{\mathrm{galaxies}} = \delta_{\mathrm{dark~matter}}\mathrm{.}
\end{equation}

However, it was realized in the early 1980s that this assumption was
quite simplistic. In particular, if galaxies form preferentially in
the regions of greatest dark matter density, one generically expects
the galaxies to be more clustered than the dark matter (Kaiser 1984;
Bardeen {\it et al.}~1986). This idea gave theorists a new free
parameter, the {\it biasing} parameter $b$, with which to fit their
models; it was particularly valuable, for example, in reconciling the
$\Omega=1$ CDM models with observations (Davis {\it et al.}~1985). (In
retrospect, the value of $b$ needed to fit the CDM model is quite
unrealistic, but that is another story). In particular, in the
appropriate limit of Kaiser's original formulation of bias, one
expects that the galaxy and mass density fields should be proportional
to one another:
\begin{equation}
\delta_{\mathrm{galaxies}} = b \,\delta_{\mathrm{dark~matter}}\mathrm{.}
\end{equation}
This linear bias model has been the default for most analyses of
large-scale structure. For example, people have used Equation (2)
({\it e.g.}, Sigad {\it et al.}~1998) or its integral equivalent ({\it
e.g.}, Willick {\it et al.}~1997) to compare redshift and peculiar
velocity data; assuming the linear bias relation, it can be phrased:
\begin{equation}
\nabla\cdot{\bf v}({\bf r}) = - \frac{\Omega^{0.6}}{b}
\delta_{\mathrm{galaxies}}({\bf r})\mathrm{.}
\end{equation}
Thus, one ends up not measuring the really interesting quantity
$\Omega$, but the somewhat awkward combination
$\beta\equiv\Omega^{0.6}/b$. 

The linear bias model is just a parameterization of our ignorance, and
is certainly over-simplistic, at least on small to moderate scales
(say, less than 20 $h^{-1}$ Mpc). It has been known for a long time
that the bias parameter must be a function of galaxy type. After all,
one sees a preponderance of spiral galaxies in the field, but they are
completely absent in the cores of rich clusters ({\it e.g.}, Dressler
1980). The large-scale distributions of galaxies of different types
are not identical. Moreover, the bias relationship must be nonlinear
at some level; the morphology-density relation in clusters shows that
the density of spirals is a non-monotonic function of the density of
ellipticals; therefore they cannot both be linear functions of the
underlying dark matter! The bias relationship is almost certainly not
completely deterministic; the detailed physics that determines whether
a galaxy forms in a given place is not a function purely of the
present-day dark matter density, and thus the galaxy density field
depends also on additional physical quantities. The bias relationship
must also be a function of smoothing scale $R$; on small scales, the
rms density fluctuations are large,
$\sigma\equiv\avg{\delta^2}^{1/2}\gg 1$, and there are regions in
which $\delta=-1$, devoid of matter. But if $b>1$ in Equation (5) (as
models would suggest), one would have the unphysical situation of
$\delta_{\mathrm{galaxies}}<-1$. Finally the bias relationship is a
function of redshift: as gravity pulls galaxies and dark matter alike
into deep potential wells, one expects $b$ to approach unity with
time. Indeed, Adelberger {\it et al.}~(1998; see also Steidel, this
conference) observe clustering at $z\approx 3$ which is as strong as
clustering today. Clustering of matter grows with time, so the
underlying dark matter distribution should be appreciably weaker back
then, implying that the effective galaxy bias was quite a bit stronger
at $z\approx 3$ than it is today.

One way to get a handle on all these complications is to turn to
cosmological simulations. In particular, Blanton {\it et al.}~(1998)
have examined the relative distribution of galaxies and dark matter in
a simulation which models both the gravitational physics of dark
matter and the gas physics of the baryons. The simulation handles star
formation by converting gas into collisionless particles in regions
with infalling gas, with cooling times below the local dynamical time,
and masses above the Jeans mass. One can then look at the relationship
between the density field of these collisionless particles (which we
take as a proxy for the galaxy density field) and the dark matter
density field.

The resulting bias relationship is nonlinear,
stochastic, and is a strong function of galaxy {\it age}.  These
properties are revealed in Figure 1, which shows as a greyscale the
conditional probability $P(1+\delta_g|1+\delta)$ and as the solid line
the conditional mean $\avg{1+\delta_g|1+\delta}$, where all quantities are
defined with a top hat filter of radius 1 $h^{-1}$ Mpc.  Each panel
shows the results at $z=0$ for galaxies formed at different epochs, as
labeled.  Note that the oldest galaxies are the cleanest tracers of
the dark matter distribution, in that the scatter around the mean
galaxy-dark matter density relation is small. However, the youngest
galaxies show a very nonlinear, even non-monotonic, relation with the
dark matter; they are underrepresented in the very densest regions of
the dark matter map (reminiscent, indeed, of spirals in the cores of
clusters, although in the real universe clusters still represent
appreciable overdensities in the distribution of late-type galaxies;
Strauss {\it et al.}~1992a). In addition, the scatter around the mean
density relation for the youngest galaxies is quite sizeable.

\begin{figure}
\vspace{3.0in}
\includegraphics{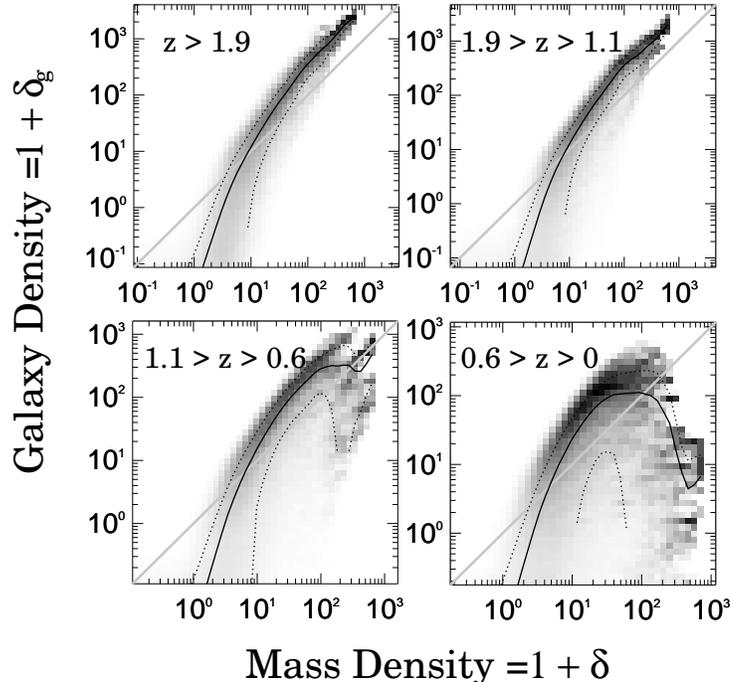}
\caption{Galaxy mass density as a function of dark matter density for
each age quartile, at 1 $h^{-1}$ Mpc radius top hat smoothing.  Each
panel lists the range of formation redshifts included. Shading is a
logarithmic stretch of the conditional probability $P(1+\delta_g |
1+\delta)$.  Solid lines indicate $\avg{1+\delta_g|1+\delta}$; dotted
lines indicate the 1$\sigma$ deviation from the mean.}
\end{figure}

In these simulations, the relationship between galaxies and mass also
depends on scale. In Figure 2, we show the bias
$b\equiv\sigma_g/\sigma$ calculated on various scales.  The obvious
scale-dependence of $b$ is due to the dependence of the galaxy
formation process on temperature. The temperature sets the local Jeans
mass, which partly determines whether star-formation occurs: the
higher the temperature, the greater the overdensity needed to form
stars. On small scales the temperature is proportional to the
gravitational potential $\phi$. Note that in Fourier space, ${\tilde
\phi}(k) \propto {\tilde \delta}(k)/k^2$. For high $k$, then, there is
little power in the potential or temperature fields; {\it i.e.}~these
fields are {\it smoother} than the density field.  Thus, 
temperature correlates over large scales; furthermore, on these large
scales it correlates with density as well.
Thus the dependence of galaxy
formation on temperature can couple the galaxy density on small scales
with the dark matter density on larger scales. As Blanton {\it et
al.}~(1998) show, this coupling causes scale-dependence of the bias
relation. The dependence of galaxy formation on local gas temperature
is likely to be important in any galaxy formation scenario; thus, this
scale-dependence may be generic.

\begin{figure}
\vspace{3.0in}
\includegraphics{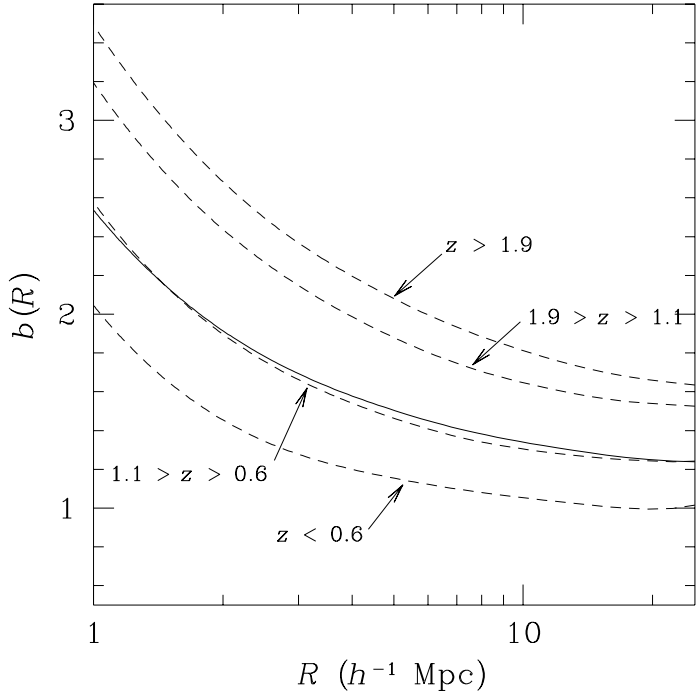}
\caption{The bias $b(R)\equiv\sigma_g(R)/\sigma(R)$, where $R$ refers to
the top hat smoothing radius. Solid
line indicates all galaxies. Dashed lines indicate each age
quartile, with range of formation redshifts listed. Note the strong
scale-dependence, and that old galaxies are more biased than
young.}
\end{figure}

The work ahead is evaluate the consequences of this rather complicated
bias relationship for interpretations of statistical measures of
large-scale structure. It is known, for example, that stochasticity in
the bias relation will systematically affect analyses comparing
peculiar velocity and density fields based on Equation (2) (Dekel \&
Lahav 1998); it may turn out that this effect is large enough to
explain the discrepancy various workers are finding in analyses of
existing datasets in the inferred value of $\beta$.

\section{The Impact of IR Astronomy on Large-Scale Structure Studies}

What is the connection of all of this to the subject of this meeting?
Infrared astronomy has had a large impact on the study of large-scale
structure, and promises to have an increasing role, as we argue in
these final remarks. In many ways, the near-infrared is an optimal
wavelength regime in which to study galaxies for statistical and
dynamical purposes. The problems with both foreground and internal
extinction are minimized, and the SEDs of the red stars that make up
the bulk of the stellar mass of galaxies peak in the near-IR. However,
it has only been relatively recently that infrared instrumentation has
advanced to the point that large-scale surveys are possible.

Thus it was realized quite early on that the near-IR was a
particularly good place to do galaxy photometry for the Tully-Fisher
relation (Aaronson, Huchra, \& Mould 1979), but the first large-scale
survey of galaxies on the sky in the near-IR, the Two-Micron All Sky
Survey, is only being carried out now (Skrutskie, this meeting). 2MASS
will be invaluable for large-scale structure studies for a number of
reasons. Having two identical telescopes observing in the two
hemispheres with the same instrumentation, with data reduced
identically, will result in a uniform galaxy catalog free from the
headaches of trying to match disparate catalogs from different regions
of the sky ({\it cf.}, Santiago {\it et al.}~1996). This, coupled with the
small effect of the Zone of Avoidance in the K Band, means that
one can study galaxy clustering on large angular scales effectively
with the 2MASS data. The galaxy catalogs generated from the {\it IRAS}
database at 60$\mu$m ({\it cf.}, Fisher {\it et al.}~1995) share many of these
virtues, and indeed have been used for a large range of large-scale
structure studies (as reviewed by Strauss \& Willick 1995), but 2MASS
will be sensitive to the early type galaxies that are mostly absent
in {\it IRAS}, and it will go appreciably deeper, with far better
sampling of the density field. In particular, the dipole moment of the
galaxy distribution can be compared with the peculiar velocity of the
Local Group to infer the depth at which large-scale flows converge;
Strauss {\it et al.}~(1992b) find hints from the {\it IRAS} data of
possible contributions to the flows on scales above 100 $h^{-1}$
Mpc. 2MASS should be able to unambiguously nail this problem. The full
analysis of large-scale structure with the 2MASS extragalactic
database will require redshifts; the survey suffers from an
embarrassment of riches, with of order $10^6$ galaxies. John Huchra is
leading a group with the ambitious goal of measuring redshifts for the
brightest 250,000 2MASS galaxies in K; this should be an absolutely
spectacular dataset for the study of large-scale structure, which will
make the {\it IRAS} redshift surveys quite obsolete.

Most analyses of large-scale structure treat all galaxies as equal
tracers of the density field. With our improved understanding of the
messy problem of biasing, and in particular a realization that the
bias of any given galaxy sample is a strong function of the way in
which they were selected, we need to include galaxy properties
explicitly in our analyses. Infrared astronomy is starting to give us
a real understanding of where the bulk of the bolometric luminosity of
galaxies is coming out; a real theme of this meeting was that SIRTF
and other tools of infrared astronomy will finally give us a clear
understanding of the full SEDs of galaxies of all different types. If
we wish to have a real understanding of the distribution of galaxies
in space, and in particular, its relation to the dark matter which
dominates the dynamics, we need to have as unbiased (in both senses of
the word!) a sample of galaxies as possible. We simply cannot learn
how to do this properly until infrared astronomy gives us the tools to
select galaxies and measure their simplest underlying physical
properties. 

\acknowledgments
This work was supported in part by NSF grant AST96-16901, the Alfred
P.~Sloan Foundation, Research Corporation, and the Princeton
University Research Board.


\begin{references}
\reference Aaronson, M., Huchra, J., \& Mould, J. 1979, ApJ, 229, 1

\reference Adelberger, K., Steidel, C., Giavalisco, M., Dickinson, M., Pettini,
M., \& Kellogg, M. 1998, ApJ, in press (astro-ph/9804236)

\reference Bardeen, J., Bond, J. R., Kaiser, N., \& Szalay, A. 1986, ApJ,
304, 15

\reference Blanton, M., Cen, R., Ostriker, J.P., \& Strauss,
M.A. 1998, ApJ, submitted (astro-ph/9807029)

\reference Colless, M. 1998, {Phil.~Trans.~R.~Soc.~Lond.~A}, in
press (astro-ph/9804079)

\reference Davis, M., Efstathiou, G., Frenk, C. S., \& White,
S. D. M. 1985, ApJ, 292, 371

\reference Dekel, A., \& Lahav, O. 1998, 
preprint, astro-ph/9806193

\reference Dressler, A. 1980, ApJ, 236, 351

\reference Fisher, K. B., Huchra, J. P., Davis, M., Strauss, M. A.,
Yahil, A., \& Schlegel, D. 1995, ApJS, 100, 69

\reference Gunn, J. E., \& Weinberg, D. H. 1995, in {\it
Wide-Field Spectroscopy and the Distant Universe}, ed.\ S. J. Maddox and A.
Arag\'on-Salamanca (Singapore: World Scientific), 3

\reference Kaiser, N. 1984, ApJ, 284, L9

\reference Santiago, B.X., Strauss, M.A., Lahav, O., Davis,
M., Dressler, A., \& Huchra, J.P. 1996, ApJ, 461, 38

\reference Schmidt, B.P. {\it et al.} 1998, ApJ, in press (astro-ph/9805200)

\reference Shectman, S.A., Landy, S.D., Oemler, A., Tucker, D.L., Lin,
H., Kirshner, R.P., \& Schechter, P.L. 1996, ApJ, 470, 172

\reference Sigad, Y., Eldar, A., Dekel, A., Strauss, M.A.,
\& Yahil, A. 1998, ApJ, 495, 516

\reference Strauss, M.A. 1998a, in {\it Formation of Structure in the
Universe}, edited by Avishai Dekel and Jeremiah P. Ostriker
(Cambridge: Cambridge University Press), $172$

\reference Strauss, M.A. 1998b, Nature, in press

\reference Strauss, M. A., Davis, M., Yahil, A., \& Huchra,
J. P. 1992a, ApJ, 385, 421

\reference Strauss, M. A., \& Willick, J. A. 1995, Phys.~Rep., 261, 271

\reference Strauss, M. A., Yahil, A., Davis, M., Huchra, J. P., \& Fisher,
K. B. 1992b, ApJ, 397, 395

\reference Willick, J.A., Strauss, M.A., Dekel, A., and
Kolatt, T. 1997, ApJ, 486, 629
\end{references}
\end{document}